\documentstyle[pre,aps,epsfig,twocolumn]{revtex}
\draft

\begin{document}

\title{Magnetization reversal times in the 2D Ising model}

\author{Kevin Brendel, G.T. Barkema and Henk van Beijeren}
\address{Theoretical Physics, Utrecht University, Leuvenlaan 4, 3584 CE
Utrecht, the Netherlands}

\date{\today}

\maketitle

\begin{abstract}
We present a theoretical framework which is generally applicable to the study
of time scales of activated processes in systems with Brownian type dynamics.
This framework is applied to a prototype system: magnetization reversal times
in the 2D Ising model. Direct simulation results for the magnetization reversal
times, spanning more than five orders of magnitude, are compared with
theoretical predictions; the two agree in most cases within 20\%.
\end{abstract}

\vspace{0.2cm}

\section{Introduction}

Activated processes that can be described with some type of Brownian dynamics
are abundant in the world around us. Well-known examples are the nucleation of
droplets in an undercooled gas or of crystals in an undercooled liquid,
chemical reactions and the escape of a protein from a misfolded state. A
prototype system to study such phenomena numerically is the well-known Ising
model. Above the so-called critical temperature, in absence of an external
magnetic field, up- and down-pointing spins are roughly equally abundant. Below
the critical temperature, the system prefers to be in either of two states: one
state with a positive magnetization in which most spins are pointing up, and
one state with a negative magnetization. As long as the system size remains
finite, reversals of the magnetization - transitions between positive and
negative magnetization - are possible and will occur at a certain average
frequency. These processes are activated, since configurations with
magnetization close to zero have a higher free energy than typical
configurations with a magnetization close to either of the equilibrium values.

In this manuscript, we study the time scales associated with magnetization
reversal. A theoretical framework is outlined which is generally applicable to
activated processes in systems with Brownian type dynamics, and compared to
high-accuracy computer simulations. From a practical point of view
magnetization reversals are also of great interest because of applications in
memory devices and the like. One wants to have rapid switching of magnetization
under reversals of an external field, but no spontaneus reversals of the
magnetization, even if the external field has been turned off. In the
literature much attention has been paid to reversal time distributions in the
presence of a driving field \cite{riknov,mischak}, but spontaneous reversals in
the absence of a field have hardly been studied. Here we consider the latter
case for the prototypical case of an Ising model with periodic boundary
conditions. We identify the leading scenario for reversals of the magnetization
and show that the process may be described to a good approximation by a
one-dimensional diffusion process over a potential barrier.

Our manuscript is organized as follows. In section~\ref{sec:model}, we describe
the model that we study in detail. Next in section~\ref{sec:theory}, we outline
the theoretical framework which is generally applicable to activated processes
in systems with Brownian type dynamics. We then apply this framework to our
prototypical model ---magnetization reversal in the Ising model. In
section~\ref{sec:sim} we compare the theoretical predictions with high-accuracy
computational results.

\section{Detailed description of the model}
\label{sec:model}

We consider the Ising model on a $B\times L$ rectangular lattice with periodic
(helical) boundary conditions, with the Hamiltonian
\begin{equation}
H=-J\sum_{\langle i,j\rangle} \sigma_i \sigma_j,
\end{equation}
in which $\sigma_i=\pm 1$ is the spin at site $i$ and $J$ is the coupling
constant. The summation runs over all pairs of nearest-neighbor sites; those of
site $i$ are $j=i\pm 1$ modulo $N$ and $j=i\pm B$ modulo $N$, with $N= BL$. The
magnetization is defined as $M \equiv \sum_i \sigma_i$; it can take values
$M=-N,-N+2,\dots,N$; all through this manuscript, we restrict ourselves to
systems in which both $B$ and $L$ are even. As a consequence, $M$ takes only
even values, and summations over a range of possible magnetizations only run
over even numbers, with an increment of 2.

A well-known property of the model is that at sufficiently low temperatures,
the distribution of the magnetization $M \equiv \sum_i \sigma_i$ becomes
bimodal: the spins in a configuration tend to align around the two preferred
values $\pm M_0$. For infinite systems this occurs at all temperatures below
the critical temperature $T_c=J/(0.44069k_B)$, for finite systems this range
starts at slightly different temperatures.

The system evolves in time according to single-spin-flip dynamics with
Metropolis acceptance probabilities~\cite{metro}. If $C_i$ is the configuration
after $i$ proposed spin flips, a trial configuration $C'_{i+1}$ is generated by
flipping a single spin at a random site. This trial configuration is then
either accepted ($C_{i+1}=C'_{i+1}$) or rejected ($C_{i+1}=C_i$); the
acceptance probability is given by
\begin{equation}
P_a= \min \left[ 1, \exp(-\beta (E(C'_{i+1})-E(C_i))\right],
\end{equation}
in which $\beta=1/(k_BT)$ with Boltzmann constant $k_B$ and temperature $T$.
The time scale is set such that in one unit of time, on average each spin is
proposed to be flipped once. So in our system, in one unit of time we perform
$BL$ Monte Carlo steps.

\begin{figure}
\epsfxsize=8cm
\epsfbox{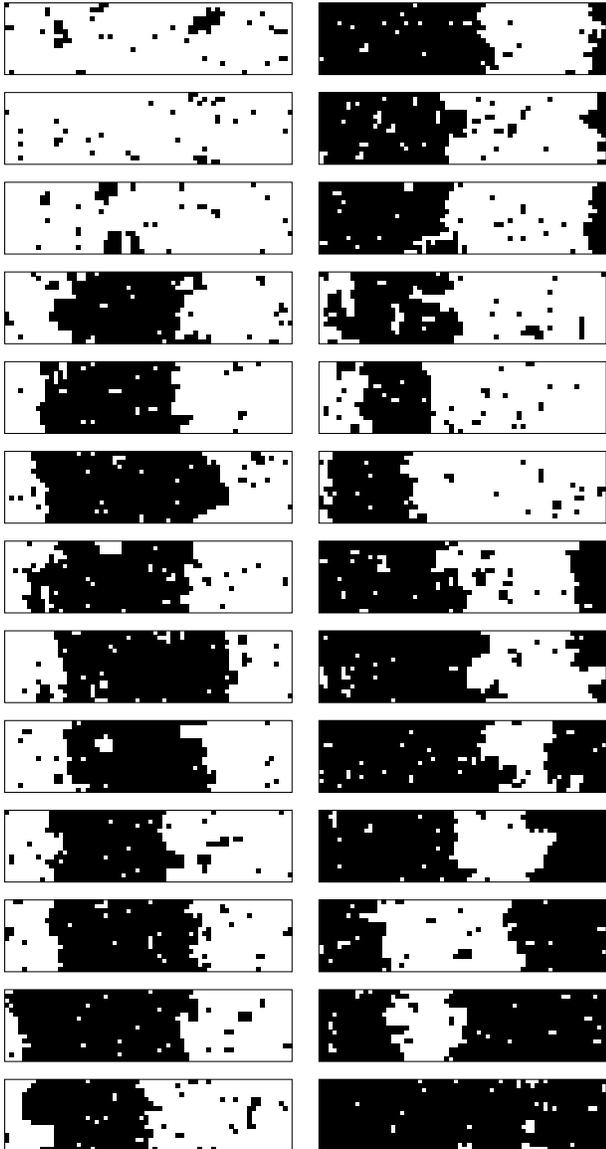}
\caption{Snapshots of the the transition from a state with most spins down to a
state with most spins up in the $16\times 64$ Ising model, taken at equal time
intervals of 500 attempted spin flips per site, at inverse temperature $\beta
J=0.5$.}
\label{fig:path1}
\end{figure}

Our current interest in the Ising model with single-spin-flip dynamics stems
from the fact that in finite systems, the configurations will occasionally
switch between states in which the magnetization is either negative or
positive, through an activated process. The dominant pathway at low
temperatures consists of the formation of a single pair of closed interfaces
in the shorter periodic direction (for $B\neq L$), which perform a relative
diffusive motion around the longer periodic direction and annihilate after
meeting each other through the periodic boundary. A series of snapshots,
illustrating this process, is presented in figure~\ref{fig:path1}.

\section{Theoretical framework}
\label{sec:theory}

To study the behavior of times between magnetization reversals at
temperatures below the critical one we may consider an ensemble of a
large number of systems prepared in states with a magnetization close
to the equilibrium value $-M_0$ and study the rate at which these
systems reach the value $M=M_0$, after which they are removed from the
ensemble. Alternatively, the behavior of times between zero-crossings
of the magnetization can be studied, as discussed below.  The spin-flip
dynamics described above may be represented by a master equation for the
probability distribution $P({\bf\sigma})$ of finding a system in the state
${\bf\sigma}$ at time $t$. Due to the huge number of possible states this
master equation cannot be solved analytically or even numerically for
system sizes of practical interest. Therefore, as an approximation we
assume that we may replace the exact master equation by an approximate
master equation for the probability $P(M,t)$ of finding a system with
magnetization $M$ at time $t$. The form of this equation is
\begin{eqnarray}
\frac{d P(M,t)}{d t} &=& \Gamma_{M,M+2}P(M+2,t)+\Gamma_{M,M-2}P(M-2,t)
\nonumber\\
&-& (\Gamma_{M+2,M}+\Gamma_{M-2,M})P(M,t),
\label{master}
\end{eqnarray}
with $\Gamma_{M',M}$ the transition rate from $M$ to $M'$. It ignores
the fact that the actual transition rates will depend on the geometry of
the state under consideration, with the idea that in typical cases spin
states of very similar geometry will dominate the set of states with
given $M$. Besides on $M$ and $\beta$ the transition rates will depend
on the geometry of the system, that is on $B$ and $L$. To estimate the
scaling behavior of these dependencies we make two further simplifying
assumptions:

\begin{itemize} 
\item First of all, requiring $B\le L$, we assume that states with $M\ne
\pm M_0$ consist of a single strip of opposite magnetization separated
by two phase boundaries of length $B$ from the majority spin phase. The
relevant changes of $M$ then will be caused by displacements of these
boundaries due to flips of spins along them. The total number of spins
available for this will be proportional to $B$, hence the transition
rates should also be proportional to $B$ and independent of $L$.

This approximation will not be valid for values of $M$ that are too close to
$\pm M_0$, as for these the opposite magnetization will typically be found in a
closed cluster rather than in a strip. It will turn out though that the
contributions from these $M$-values to the reversal frequency are very small
for systems of reasonable length. Therefore using the approximation $\Gamma
\sim B$ also here does not harm. Further we neglect the possibility of having
more than one strip of opposite magnetization. When the temperature gets close
to the critical one, when $B$ becomes small or when $L$ becomes very large,
this may not be a good approximation. We will come back to this in section
\ref{sec:disc}. In our further theoretical treatment we will assume that we are
in a situation where the single-strip approximation is justified.

\item Further we neglect changes due to fluctuations of the magnetization
caused by the growth and shrinkage of small clusters of the minority
spin type.  On average these do not contribute to magnetization reversals,
so this approximation should be allowed. In case one considers the first
passage frequency through zero magnetization there is a small effect,
to which we will come back in section \ref{sec:disc} too.

\end{itemize}

In order that the equilibrium distribution be a stationary solution of
the master equation we impose the condition of detailed balance
\begin{equation}
\frac{\Gamma_{M,M+2}}{\Gamma_{M+2,M}}=\exp\left[ \beta(F(M)-F(M+2))\right],
\label{detbal}
\end{equation}
where $\beta F(M)=-\ln P_{eq}(M)$, with $P_{eq}(M)$ the equilibrium
probability of magnetization $M$. In section \ref{sec:3B} it will be
explained how this condition may be combined with simulation results for
interface diffusion to obtain approximations for all transition rates
that lead to a good overall prediction of the reversal times.

The long-time reversal frequency as predicted by the master equation
(\ref{master}) follows as the largest eigenvalue of this equation, supplemented
with an absorbing boundary at $M=A$, with either $A=M_0$ or $A=0$. The first
choice corresponds to a real reversal of the magnetization, the second one to a
first return to the value $M=0$. After this the system will have equal
probabilities to actually reverse its magnetization or to return to the
equilibrium magnetization value it came from, hence this return frequency
should be twice the reversal frequency. The absorbing boundary condition is
implemented by setting $\Gamma_{A-2,A}$ equal to zero.

The largest eigenvalue $-\nu$ of $\Gamma_{M,M'}$ in eq. (\ref{master}),
as well as the corresponding eigenvector $P_0(M)$, may be found by
requiring that the net current away from magnetization $M$ assumes the
value $\nu P_0(M)$. Using conservation of probability one easily checks
that this may be expressed as
\begin{eqnarray}
\Gamma_{M+2,M}P_0(M) &-& \Gamma_{M,M+2}P_0(M+2)=\nu\sum_{m\le M} P_0(m),
\label{Gamma} \\
\nu &=& \frac{\Gamma_{A,A-2} P_0(A-2)}{\sum_{m\le A-2}P_0(m)}.
\end{eqnarray}

In eq. (\ref{Gamma}) $P_0(m)$ on the right hand side may be approximated, up to
a normalization factor, by $\exp(-\beta F(m))$, because the sum is dominated by
the terms with small $m$-values, for which this approximation is excellent.
This one may check in hindsight against the solution obtained. With this
approximation the equation may be solved recursively for $P_0(M)$ in terms of
$P_0(A-2)$ for $M=A-4,A-6,...$, with the result
\begin{eqnarray}
P_0(M) &=& \sum_{M\le m \le A-2}\frac{\Gamma_{A,A-2}}{\Gamma_{m+2,m}}\exp\left[
\beta(F(m)-F(M))
\right] \nonumber\\
   &\times & \frac {\sum_{n\le m}\exp\left[ -\beta F(n)\right] }
                   {\sum_{n'\le A-2}\exp\left[ -\beta F(n')\right]}
P_0(A-2).
\label{eq:mode}
\end{eqnarray}
The summations over $m$ and $M$ are dominated by values of $m$ close to zero,
combined with values for $M$ close to $-M_0$, for which the sum over $n$ is
basically independent of $m$. Therefore the reversal frequency may be obtained
as
\begin{eqnarray}
\nu & = & \frac{\Gamma_{A,A-2}P_0(A-2)}{\sum_{M\le A-2} P_0(M)} \nonumber\\
    & = &\left(\sum_{m=-M_0}^{A-2} \frac{\exp\left[ \beta F(m)\right]}
	{\Gamma_{m+2,m}} \sum_{n\le A-2}
	\exp\left[ -\beta F(n)\right]\right)^{-1}.
\label{seven}
\end{eqnarray}
The restriction of the summation over $m$ to values \mbox{$m>-M_0$} is needed
to avoid large spurious contributions from $m<-M_0$. The result in eq.
(\ref{seven}) is well-known. It is usually derived by considering a state with
a stationary current in which mass is inserted at a constant rate on one side
(e.g.\ at $M=-B\times L$ in our case) and taken out as soon as it reaches the
absorbing boundary (see e.g.\ \cite{hanggi}, section IV E). In that case the
replacement of the sum over $n$ by a constant is exact.

\section{Simulations and results}
\label{sec:sim}

In order to apply the above theoretical framework to magnetization
reversal times in the Ising model, the two ingredients required are:
(i) the equilibrium probability $P_{eq}(M)$ to find the system in a state
with magnetization $M$; and (ii) the transition rates $\Gamma_{M',M}$
from magnetization $M$ to $M'$. We obtain these two ingredients via two
different computational approaches.

\subsection{free energy landscape}

\begin{figure}
\epsfxsize=5cm
\epsfbox{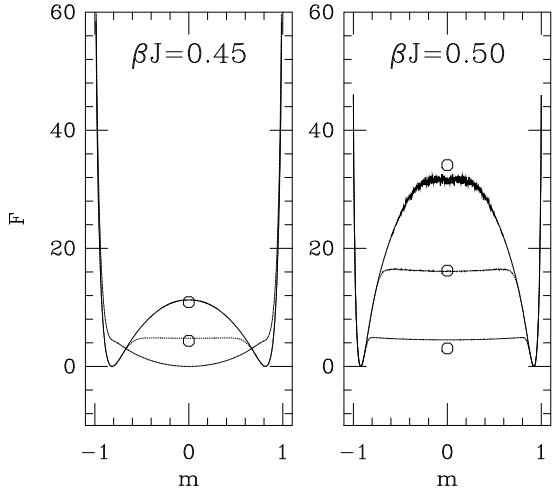}
\caption{Free energy as a function of magnetization at inverse temperatures
$\beta J=0.45$ and $\beta J=0.50$. The number of spins is kept constant at
1024. The solid lines represent the $32\times 32$ system, the dotted lines
represent the $16\times 64$ system, and the dash-dotted lines represent the
$8\times 128$ system. The free energies at $M=0$ corresponding to equation
(\ref{eq:zeroprob}) are also indicated. A constant is added to the free energy
curves, such that the minimal value of $F(M)$ is zero in all cases.
\label{fig:fhisto}}
\end{figure}

For various values of $L$ and $B$ and various temperatures we make histograms
of the distribution of the magnetization $M=\sum_i \sigma_i$ and the energy.
Since the probability that a state with energy $E$ occurs at an inverse
temperature $\beta$ is proportional to the Boltzmann weight $\exp(-\beta E)$, a
histogram made at a certain temperature provides information about the
probability distribution at nearby temperatures as well. We use the multiple
histogram method~\cite{ferrenberg} to combine the information from simulations
at various temperatures. In this way, histograms for the magnetization over a
wide range of temperatures can be obtained. As defined above, the free energy
$F(M)$ is related to the probability that a certain magnetization $M$ occurs by
$P_{eq}(M)=e^{-\beta F(M)}$. The method of determining free energies of
configurations constrained to some value of a coordinate along a pathway
through phase space has been used extensively, for instance by Auer and
Frenkel~\cite{frenkel}. In figure \ref{fig:fhisto} the free energy as a
function of the magnetization per spin $m=M/N$ is plotted at two different
temperatures and three different values of $L/B$.

In order to compare these results with theory we note that the probability
$P_{eq}(0)$ of finding the system in a state with $M=0$ can be obtained
from eq. (4.23) of chapter V of \cite{mccoywu} as follows. First note
that this equation gives the following approximation for the partition
function of a periodic anti-ferromagnetic system with one phase boundary
\cite{mwremark}, which is equivalent to a ferromagnetic system with an
anti-periodic boundary:
\begin{equation}
Z_1=\frac{Z_0}{2} \sum_{k=1}^{2L}
  \left( \frac{A(k)-\sqrt{A^2(k)-c^2}}{A(k)+\sqrt{A^2(k)-c^2}} \right)^{B/2},
\label{onsager}
\end{equation}
where
\begin{eqnarray*}
A(k) & = & (1+z^2)^2-2z(1-z^2) \cos\frac{\pi k}{L}, \\
c    & = & 2z(1-z^2),  \\
z    & = & \tanh\beta J,
\end{eqnarray*}
and $Z_0$ the partition function for a homogeneous system (cf. (4.20) of
\cite{mccoywu}). Neglecting the possible interaction between two phase
boundaries (or interfaces) we conclude that the partition function $Z_2$ for a
system with two interfaces is given by $Z_2/Z_0=(1/2)(Z_1/Z_0)^2$. The factor
$1/2$ arises because otherwise we would count each configuration twice:
interchanging the locations of the two interfaces does not give a different
configuration. If we leave the positions of the two interfaces completely free
the magnetization is distributed almost uniformly over all possible even values
between $-M_0$ and $M_0$. Therefore, fixing $M$ to 0 leads to a reduction of
the partition function by a factor of $M_0$. Thus we arrive at the following
result for $P_{eq}(0)$:
\begin{eqnarray}
P_{eq}(0) & = & \frac{Z_2}{M_0Z_0} \nonumber \\
     & = & \frac{1}{8M_0} \left[ \sum_{k=1}^{2L} \left(
           \frac{A(k)-\sqrt{A^2(k)-c^2}}{A(k)+\sqrt{A^2(k)-c^2}}
	   \right)^{B/2} \right]^2.
\label{eq:zeroprob}
\end{eqnarray}
For square systems this result has to be modified, because there are two ways to
make a strip with opposite magnetization: the interfaces can lie in the
horizontal as well as in the vertical direction. This gives an additional factor
2 in the equation for $P_{eq}(0)$.

\subsection{interface diffusion coefficient} \label{sec:3B}

The second ingredient for the theoretical framework in section~\ref{sec:theory}
consists of the transition rates $\Gamma_{M',M}$ from magnetization $M$ to
$M'$. As discussed there, we are actually mostly interested in the contribution
to the transition rates that arises from the diffusion parallel to the longer
periodic direction of interfaces that span the shorter one. To estimate this
diffusion coefficient numerically, we study systems with anti-periodic boundary
conditions: the spins in two neighboring sites $i<N-1$ and $j=i+1$, or in
$i<N-B$ and $j=i+B$ are aligned if $\sigma_i=\sigma_j$, whereas the spins in
two neighboring sites $i=N-1$ and $j=0$, or in $i\ge N-B$ and $j=i+B-N$,
are aligned if $\sigma_i=-\sigma_j$.

The interface location $x(0)$ is initially defined as the magnetization $M(0)$.
As long as $M\in [-0.8N,0.8N]$, steps $\Delta x$ in the interface location are
equal to changes $\Delta M$ in the magnetization, {\it i.e.}, $x(t+\Delta
t)-x(t)=M(t+\Delta t)-M(t)$. Once the interface gets close to the anti-periodic
boundary, the magnetization does not uniquely determine the interface location.
As soon as $M$ is no longer in the interval $[-0.8N,0.8N]$, we shift the
anti-periodic boundary away from the interface over half the system size (which
will bring the magnetization back in this range) and then continue.

In practice, we achieve this by switching from monitoring $M$ to monitoring
$M'=\sum_{i=0}^{N/2-1}\sigma_i - \sum_{i=N/2}^{N-1}\sigma_i$, and $\left|
x(t+\Delta t)-x(t)\right| = \left| M'(t+\Delta t)-M'(t)\right|$. The sign of
the steps in $x(t)$ depends on the location of the interface: in the lower half
this sign will be unchanged, whereas in the upper half it will be reversed. As
soon as $M'$ leaves the interval $[-0.8N,0.8N]$ we switch back to measuring
steps in the original magnetization $M$. Also here, the sign of the steps in
$x(t)$ depends on the location of the interface. Note that $x$ is not confined
to the interval $[-N,N]$. The diffusion coefficient $D$ is then obtained from
the time-dependent interface location $x(t)$ as
\begin{equation}
D=\lim_{t\rightarrow \infty}
\left[\frac{\langle(x(t)-x(0))^2\rangle}{2t}\right]. \label{eq:D}
\end{equation}
We expect that the interface diffusion is independent of $L$, and that it grows
linearly with $B$ since the number of sites where a spin flip moves the
interface grows linearly with its width; however, for small $B$ corrections
arise due to the periodicity (or helicity) of the boundaries. Consequently, we
expect
\begin{equation}
D(B,L,\beta J)=g(\beta J) B+c. \label{eq:g}
\end{equation}

\begin{figure}
\epsfxsize=8cm
\epsfbox{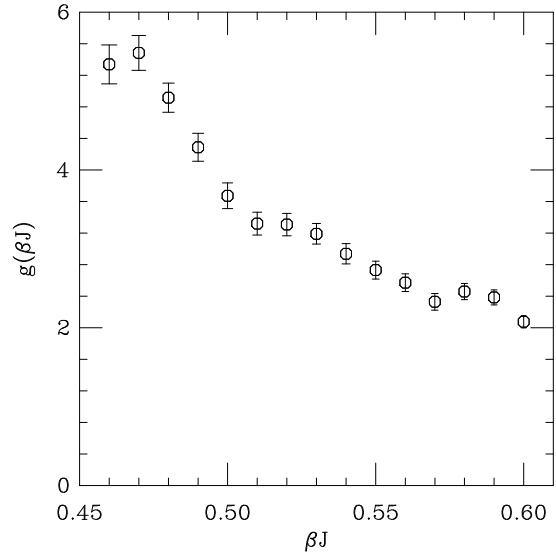}
\caption{Monte carlo measurements of the diffusion coefficient per
interface length $g(\beta J)$, as a function of inverse temperature
$\beta J$.
\label{fig:g}}
\end{figure}

For temperatures close to $T_c$ and small $B$, the system might occasionally
contain more than a single interface. In that case, equation (\ref{eq:D}) would
overestimate the diffusion coefficient of an interface. Since the additional
free energy cost of interfaces increases linearly with $B$, this unwanted
contribution to $D$ decreases exponentially with $B$. Taking this into account,
our approach is to measure for various values of $\beta J$, $B$, and $L$ the
diffusion coefficient $D$ via eq. (\ref{eq:D}). Next we determine the function
$g(\beta J)$ in equation (\ref{eq:g}); the results are plotted in figure
\ref{fig:g}. Indeed we find that for increasing $B$ and at temperatures not too
close to $T_c$ the diffusion coefficient rapidly becomes independent of $L$
within the ranges of $L$-values we considered. In our theoretical framework we
then use as approximation for the jump rates
\begin{equation}
\min[\Gamma_{M,M+2},\Gamma_{M+2,M}]=\frac{g(\beta J) B}{2}.
\label{jumprates}
\end{equation}
This may be understood from the observations that on the one hand one has
$D=4\Gamma$, because the jumps in magnetization go by units of 2, on the other
hand $D$ satisfies (\ref{eq:g}) with $B$ replaced by $2B$ because there are
two interfaces. The above equation, in combination with eq. (\ref{detbal}) and
the free energy as a function of magnetization, specifies all the transition
rates. Using (\ref{seven}) we can now predict the magnetization reversal times.
The results are shown in Table~\ref{tab:estim-rev}.

\begin{table*}
\caption{Estimated values for the magnetization reversal times.
\label{tab:estim-rev}}
\begin{tabular}{llll}
$\beta J$ & $32\times 32$        & $16\times 64$   & $16\times 32$    \\
\hline
0.46      &    $1.66\cdot 10^5$ & $1.74\cdot 10^4$ & $1.19\cdot 10^4$ \\
0.47      &    $1.12\cdot 10^6$ & $4.21\cdot 10^4$ & $3.15\cdot 10^4$ \\
0.48      &    $8.88\cdot 10^6$ & $1.25\cdot 10^5$ & $9.96\cdot 10^4$ \\
0.49      &    $7.38\cdot 10^7$ & $3.97\cdot 10^5$ & $3.31\cdot 10^5$ \\
0.50      &    $6.49\cdot 10^8$ & $1.31\cdot 10^6$ & $1.12\cdot 10^6$ \\
0.51      &    $5.83\cdot 10^9$ & $4.19\cdot 10^6$ & $3.61\cdot 10^6$ \\
0.52      & $5.31\cdot 10^{10}$ & $1.23\cdot 10^7$ & $1.05\cdot 10^7$ \\
0.53      & $5.77\cdot 10^{11}$ & $3.78\cdot 10^7$ & $3.11\cdot 10^7$ \\
0.54      & $7.69\cdot 10^{12}$ & $1.23\cdot 10^8$ & $9.60\cdot 10^7$ \\
\hline
$\beta J$ & $16\times 16$       & $8\times 64$     & $8\times 32$     \\
\hline
\hline
0.46      &    $4.43\cdot 10^3$ & $2.76\cdot 10^3$ & $1.24\cdot 10^3$ \\
0.47      &    $1.06\cdot 10^4$ & $3.26\cdot 10^3$ & $1.73\cdot 10^3$ \\
0.48      &    $3.09\cdot 10^4$ & $4.72\cdot 10^3$ & $2.88\cdot 10^3$ \\
0.49      &    $9.56\cdot 10^4$ & $7.45\cdot 10^3$ & $5.11\cdot 10^3$ \\
0.50      &    $3.09\cdot 10^5$ & $1.26\cdot 10^4$ & $9.47\cdot 10^3$ \\
0.51      &    $9.60\cdot 10^5$ & $2.12\cdot 10^4$ & $1.69\cdot 10^4$ \\
0.52      &    $2.73\cdot 10^6$ & $3.32\cdot 10^4$ & $2.77\cdot 10^4$ \\
0.53      &    $8.06\cdot 10^6$ & $5.50\cdot 10^4$ & $4.71\cdot 10^4$ \\
0.54      &    $2.50\cdot 10^7$ & $9.69\cdot 10^4$ & $8.43\cdot 10^4$ \\
0.55      &    $7.64\cdot 10^7$ & $1.70\cdot 10^5$ & $1.49\cdot 10^5$ \\
0.56      &    $2.30\cdot 10^8$ & $2.96\cdot 10^5$ & $2.61\cdot 10^5$ \\
0.57      &    $7.20\cdot 10^8$ & $5.38\cdot 10^5$ & $4.75\cdot 10^5$ \\
0.58      &    $1.92\cdot 10^9$ & $8.37\cdot 10^5$ & $7.39\cdot 10^5$ \\
0.59      &    $5.54\cdot 10^9$ & $1.42\cdot 10^6$ & $1.25\cdot 10^6$ \\
0.60      & $1.77\cdot 10^{10}$ & $2.66\cdot 10^6$ & $2.34\cdot 10^6$ \\
\end{tabular}
\end{table*}

\subsection{magnetization reversal times} \label{sec:3C}

We measure the magnetization every $N$ attempted spin flips. We look for events
where the magnetization crosses from positive (more than half the spins up) to
negative or vice versa between two consecutive measurements.

\begin{figure}
\epsfxsize=8cm
\epsfbox{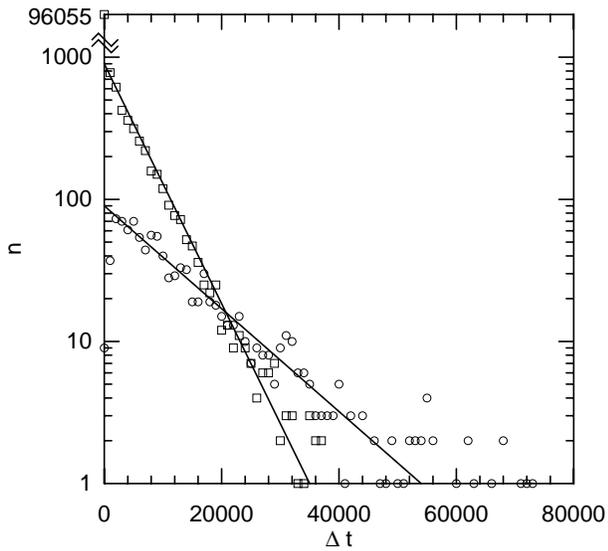}
\caption{Histogram of the times between zero-crossings of the
magnetization (squares), and the times between first occurrence of a free energy
minimum (circles) in the $16\times 64$ Ising model at $\beta J=0.45$. The solid
lines depict the fit to the data, obtained as described below.
\label{fig:tau}}
\end{figure}

We make histograms of the times between two occurrences of magnetization
reversal events. The histogram obtained in a system containing $16\times
64$ spins, at the temperature corresponding to $\beta J=0.45$, is shown
in figure~\ref{fig:tau}. For comparison we also show a histogram of the
times measured between the first time the system reaches a free energy
minimum, and the first time after this it reaches the other minimum.
We will come back to this in our discussion.

The figure shows that at long times the decay function $f(t)$ behaves as
$f(t)\sim\exp(-t/\tau)$. Here, we focus on the long-time behavior, and are
specifically interested in the escape time $\tau$. We obtain this quantity via
a fitting procedure, in which we ignore the data up to a time $t_0$, chosen
such that $f(t)$ shows exponential time behavior for $t>t_0$. Then we determine
the time $t'$ at which half of the remaining events have taken place. The
escape time $\tau$ is then obtained from $\tau=(t'-t_0)/\ln(2)$. Instead we
could have made a linear fit of all the data points beyond $t_0$, but this
makes no significant difference.

\begin{figure}
\epsfxsize=8cm
\epsfbox{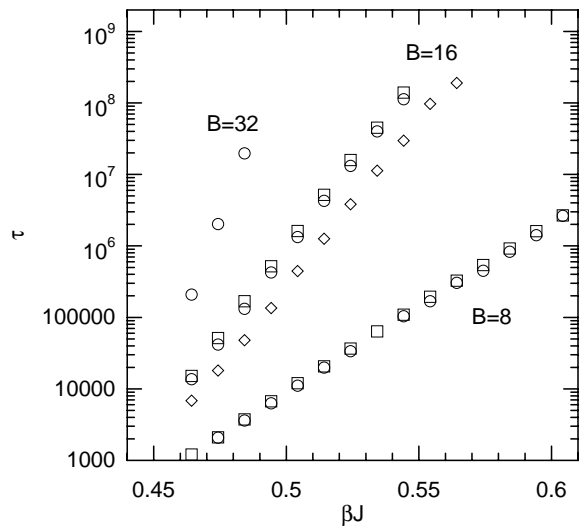}
\caption{Magnetization reversal times $\tau$ as a function of inverse
temperature $\beta J$, for various system sizes. The diamonds represent
systems with $L=16$, the circles represent systems with $L=32$, and the squares
represent systems with $L=64$.
\label{fig:revfig}}
\end{figure}

The resulting reversal times, for several system sizes and inverse
temperatures, are presented in Table~\ref{tab:direct-rev} and plotted in
figure \ref{fig:revfig}.

\begin{table*}
\caption{Directly measured values for the magnetization reversal times.
\label{tab:direct-rev}}
\begin{tabular}{llll}
$\beta J$ & $32\times 32$	& $16\times 64$       & $16\times 32$	    \\
\hline
0.46      & $2.09(6)\cdot 10^5$ & $1.52(3)\cdot 10^4$ & $1.37(1)\cdot 10^4$ \\
0.47      & $2.02(6)\cdot 10^6$ & $5.14(5)\cdot 10^4$ & $4.18(4)\cdot 10^4$ \\
0.48      & $1.97(6)\cdot 10^7$ & $1.68(1)\cdot 10^5$ & $1.32(1)\cdot 10^5$ \\
0.49      &			& $5.20(5)\cdot 10^5$ & $4.26(6)\cdot 10^5$ \\
0.50      &			& $1.62(5)\cdot 10^6$ & $1.33(2)\cdot 10^6$ \\
0.51      &			&  $5.2(1)\cdot 10^6$ & $4.28(6)\cdot 10^6$ \\
0.52      &			& $1.59(5)\cdot 10^7$ & $1.32(3)\cdot 10^7$ \\
0.53      &			&  $4.5(2)\cdot 10^7$ &  $4.0(1)\cdot 10^7$ \\
0.54      &			& $1.40(6)\cdot 10^8$ & $1.13(4)\cdot 10^8$ \\
\hline
	  & $16\times 16$	& $8\times 64$        & $8\times 32$	    \\
\hline
0.46 	  & $6.83(7)\cdot 10^3$ & $1.21(1)\cdot 10^3$ & $1.23(1)\cdot 10^3$ \\
0.47      & $1.81(2)\cdot 10^4$ & $2.10(2)\cdot 10^3$ & $2.08(2)\cdot 10^3$ \\
0.48 	  &  $4.8(2)\cdot 10^4$ & $3.75(3)\cdot 10^3$ & $3.61(4)\cdot 10^3$ \\
0.49 	  & $1.35(4)\cdot 10^5$ & $6.70(7)\cdot 10^3$ & $6.30(6)\cdot 10^3$ \\
0.50 	  & $4.45(4)\cdot 10^5$ & $1.20(1)\cdot 10^4$ & $1.11(1)\cdot 10^4$ \\
0.51 	  & $1.26(1)\cdot 10^6$ & $2.07(2)\cdot 10^4$ & $1.99(2)\cdot 10^4$ \\
0.52 	  & $3.82(5)\cdot 10^6$ & $3.66(3)\cdot 10^4$ & $3.37(3)\cdot 10^4$ \\
0.53 	  & $1.13(2)\cdot 10^7$ & $6.36(6)\cdot 10^4$ & $5.87(5)\cdot 10^4$ \\
0.54 	  & $2.97(6)\cdot 10^7$ & $1.09(1)\cdot 10^5$ & $1.04(2)\cdot 10^5$ \\
0.55 	  &  $9.7(3)\cdot 10^7$ & $1.95(3)\cdot 10^5$ & $1.68(3)\cdot 10^5$ \\
0.56 	  &  $1.9(2)\cdot 10^8$ & $3.26(6)\cdot 10^5$ & $3.05(9)\cdot 10^5$ \\
0.57 	  &			& $5.40(8)\cdot 10^5$ &  $4.5(1)\cdot 10^5$ \\
0.58 	  &			&  $9.2(2)\cdot 10^5$ &  $8.3(2)\cdot 10^5$ \\
0.59 	  &			& $1.61(2)\cdot 10^6$ & $1.41(3)\cdot 10^6$ \\
0.60 	  &			& $2.67(5)\cdot 10^6$ & $2.64(8)\cdot 10^6$ \\
\end{tabular}
\end{table*}

One sees by comparing the results of Tables I and II that in most cases the two
agree within 20\% for temperatures not too close to the critical temperature.

\begin{figure}
\epsfxsize=8cm
\epsfbox{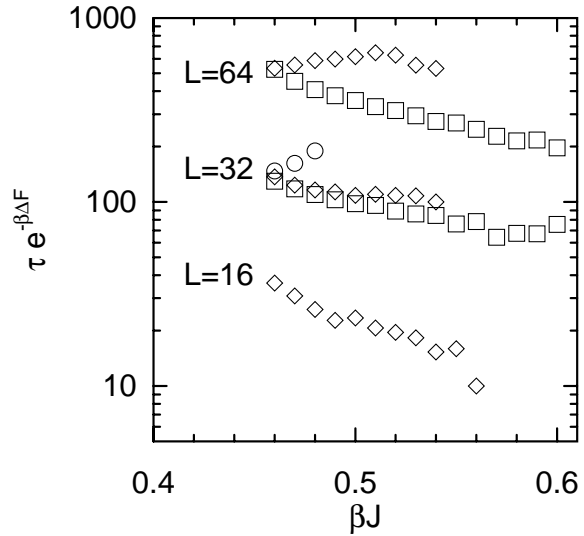}
\caption{Ratio of the magnetization reversal times $\tau$ and $\exp(\beta
\Delta F)$, as a function of inverse temperature $\beta J$, for various system
sizes. The squares represent systems with $B=8$, the diamonds represent systems
with $B=16$, and the circles represent systems with $B=32$.
\label{fig:pref}}
\end{figure}

A commonly used first approximation to the description of the time scales of
activated processes is Arrhenius' law, which states that the typical time scale
$\tau$ increases exponentially with the height of the (free) energy barrier
$\Delta F\equiv F(M=0)-F(M=M_0)$, with a prefactor $f$ that depends in a mild
way on temperature and system size,
\begin{equation}
\tau = f(\beta,B,L) \exp(\beta \Delta F).
\end{equation}
If we straightforwardly use the free energy $F(M)$ as defined before we
can obtain the heights of the free-energy barriers directly from the
histograms of the magnetization distribution. To check the accuracy
of this, we have plotted in figure \ref{fig:pref} the ratio of $\tau$
and $\exp(\beta \Delta F)$ as a function of inverse temperature $\beta
J$, for various system sizes. Clearly, if the prefactor $f$ is assumed
to be constant, this simple approximation fails to even predict the
magnetization reversal times within an order of magnitude.

A little thought reveals that, in the present case, this way of
determining the free energy barrier is not entirely satisfactory. The
reason is that for magnetization values around $M=M_0$ the free
energy increases with system size because the range of values through
which $M$ typically fluctuates, increases as the square root of system
size. As a consequence the probability of finding any specific $M$-value
decreases. Around $M=0$ on the other hand, no similar effect occurs, due
to the wide plateau of the free energy as function of $M$.  As a result
of this the decrease of probability of finding a given magnetization
value for given relative position of the two interfaces is precisely
compensated by the probabilities of finding this same magnetization value
at different relative positions. Therefore it makes more sense in this
case to define the Arrhenius factor in an alternative way as $\exp(\beta
F(M=0))$. With this definition, where again $\beta F(M)=-\ln (P_{eq}(M))$,
it may be interpreted indeed as the equilibrium probability of finding
two interfaces dividing the system into equal areas. But please notice
that if, instead of a plateau, the free energy has a maximum of a width
small compared to the range of magnetization fluctuations, the first
definition of the Arrhenius factor is to be preferred.

With the second definition the expression for the decay time in terms of the 
Arrhenius factor becomes
\begin{equation}
\tau = f'(\beta,B,L) \exp(\beta F(M=0)),
\end{equation}
in which again one hopes that the prefactor $f'$ depends on temperature and
system dimensions only in a mild way. On the basis of (\ref{seven}) and 
(\ref{jumprates}) we may conclude that for large systems $f'$ should simply be 
proportional to $L$ and independent of $B$.

\begin{figure}
\epsfxsize=8cm
\epsfbox{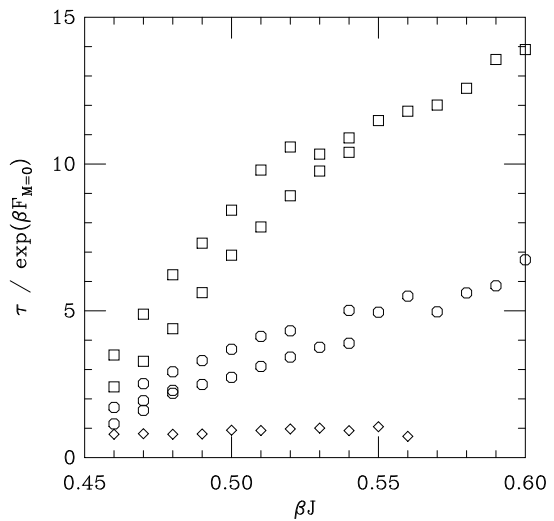}
\caption{Ratio of the magnetization reversal times $\tau$ and $\exp(\beta
F(0))\equiv P_{eq}(M=0)^{-1}$, as a function of inverse temperature $\beta
J$, for various system sizes. The squares represent systems with $L=64$,
the circles represent systems with $L=32$, and the diamonds represent
systems with $L=16$.
\label{fig:fig7}}
\end{figure}

Figure \ref{fig:fig7} shows $f'$ as a function of inverse temperature for
several system sizes. For $L=32$ and 64 both the approximate independence of
$B$ and the proportionality to $L$ are quite well-confirmed, especially if one
takes into account that the plateau widths for these system sizes are notably
less than $L$ (see figure 2). For $L=16$ the plateau becomes so narrow that
the above predictions do not apply.

\subsection{most slowly decaying mode}

\begin{figure}
\epsfxsize=8cm
\epsfbox{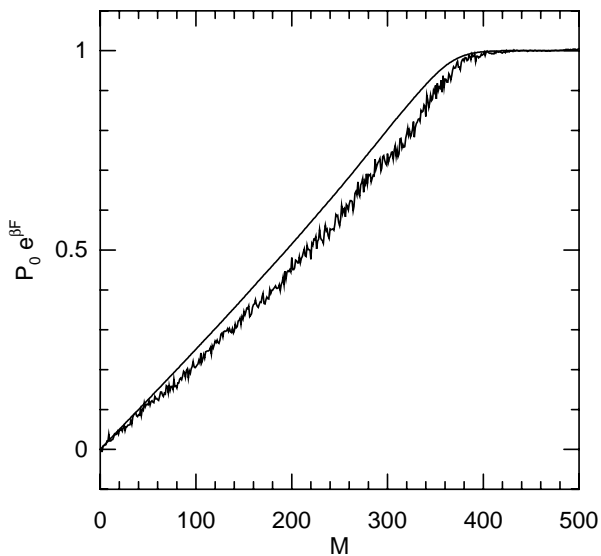}
\caption{The most slowly decaying eigenmode $P_0(M)$ divided by the equilibrium
distribution $\exp(-\beta F(M))$ for the $16\times 64$ system at $\beta J=0.5$,
obtained in two different ways as described in the text. Since this function is
antisymmetric in $M$, the left hand side of the figure $(M<0)$ has been left
out for clarity.
\label{fig:antisym}}
\end{figure}

To obtain an estimate for the magnetization reversal times, we first estimated
the most slowly decaying mode in eq. (\ref{eq:mode}). Besides comparing the
magnetization reversal times, we can also obtain the most slowly decaying
eigenmode $P_0(M)$ directly from our Monte Carlo simulations by measuring the
probability difference between the occurrence of a certain magnetization $M$
and the occurrence of the opposite magnetization $-M$, averaged over a time
scale comparable to the magnetization reversal time $\tau$. Since $P_0$ decays
much slower than the other antisymmetric (in $M$) modes, it will give the
dominant contribution to this probability difference. Figure~\ref{fig:antisym}
compares the most slowly decaying eigenmode as obtained with eq.
(\ref{eq:mode}) with the direct Monte Carlo measurements, for the $16\times 64$
system at $\beta J=0.5$.

\section{Discussion}
\label{sec:disc}

Our simulations confirm the global picture we sketched for the process
of magnetization reversals in the Ising model with stochastic dynamics:
a large cluster of opposite magnetization, originating through a
fluctuation develops into a pair of interfaces. These interfaces diffuse
around the system and annihilate, leaving the system in the oppositely
magnetized phase. Quantitatively this process may be described to a good
approximation as a diffusion process in a one-dimensional space with a
coordinate describing the total magnetization in the system.

A few remarks should be made here.
\begin{itemize}
\item First of all it may seem remarkable that in almost all cases our
theoretical prediction gives a shorter reversal time than the simulations.
Since the theory neglects processes that enhance the reversal frequency, such
as the formation of more than two simultaneous interfaces and spurious passages
of $M=0$ (see below), one might expect an overestimate of the reversal time
rather than an underestimate. The explanation of the latter comes primarily
from that part of the process in which a growing cluster of opposite
magnetisation transforms into a band around the cylinder. As the cluster will
at first be typically circular in shape (if the temperature is not too low) it
has to deform into more elliptical shape, with a longer interface than the
band, before the latter can be formed. In figure 2 the signs of this are very
clear in the form of little shoulders on the sides of the plateaus in the free
energies. The transition rates $\Gamma_{M',M}$ for the formation of these
elliptical shapes will not be proportional to $B$ but much smaller, since this
growth mainly occurs along the short sides of the cluster. The contributions
from these terms in the denominator of (\ref{seven}) are therefore larger than
estimated. Numerically they are important as long as the plateau in the free
energy is not too wide. This then will give rise to a notable decrease of the
reversal frequencies from our estimated values. Correcting for this in the
theoretical expression would require better estimates of $\Gamma_{M',M}$ in the
free energy shoulders, which seem fairly complicated to obtain. Also the 
growth rates in the magnetization regions beyond the shoulders, where the
growing clusters are even smaller, are overestimated by (\ref{jumprates}), but
these are weighted less as the free energies there are smaller.

\item The above considerations clearly reveal the possibility that the
coordinate parametrizing the "reaction path" (here the magnetization) changes
non-monotonically along this reaction path; one could imagine that the growing 
cluster, before turning into a strip, typically decreases in size for a while.
In reality this does not seem to happen, but if this were the case, parts of
the  reaction path with different cluster shape but equal magnetization would
be  lumped together, and the diffusion process along the cluster size
coordinate  would no longer correspond to the actual path taken by the cluster.
Obviously in such a case the reaction coordinate has to be redefined in such a
way that the new coordinate is monotonic along the reaction path indeed. But in
complicated situations it may not always be clear what is a proper choice for
such a coordinate.

\item
As a consequence of magnetization fluctuations in the bulk, passages of $M=0$
will be registered typically already a short while before the area between the
interfaces reaches the value $LB/2$ corresponding to an equal division of the
system between areas of positive and negative magnetization. The reason is that
the typical time scale for fluctuations of the bulk magnetization is much
shorter than that for interface diffusion, so for each location of the
interfaces the whole range of accessible magnetization values typically will be
scanned. Now in most cases this just will give rise to a negligible shift of
the time at which the situation of equal areas (this is the physically relevant
criterion) is reached, but occasionally it may happen that $M=0$ occurs, but
the system returns to the pure state it came from without ever reaching equal
areas. To check the importance of these events we may compare the average first
passage time from $-M_0$ to $M_0$ to that from $\pm M_0$ to $M=0$. If the
effect of magnetization fluctuations is negligible the ratio of these should be
2, otherwise it ought to be larger. Note that for the systems considered here
the effect of magnetization fluctuations on the first passage time from $-M_0$
to $M_0$ is much smaller than that on the first passage time from $-M_0$ to
$M=0$, because the probability of returning to $-M_0$ without reaching the pure
state of positive magnetization, once $M=M_0$ has been reached, is extremely
small. We have measured this ratio for several system sizes and temperatures,
excluding those cases where the occurrence of multiple interfaces is likely,
and found that the mean value is $1.90$, with a standard deviation of $0.15$.

\item It is of interest to investigate how the reversal times depend on the
system size parameters. From (\ref{seven}) we may conclude that for long
systems (but not so long that multiple interface pairs will occur frequently)
the reversal time will become independent of the system length $L$. For this
one should notice first of all that for $M$-values on the plateau $\exp(\beta
F(M))$ is proportional to $1/L$, because the number of ways a pair of
interfaces may be placed such that the average magnetization equals $M$, is
proportional to $L$. On the other hand the summation over $M$, which is
dominated by $M$-values on the plateau, gives rise to a factor close to $L$.
Hence to first approximation the reversal frequency $\nu$ is independent of
$L$. Remarkably this independence of $L$ in fact is observed even better by the
numerical results than by the theoretical estimates.

In section \ref{sec:3C} it was noted already that the product $f'$ of the 
reversal frequency and the Arrhenius factor $\exp(-\beta F(0))$ depends on $L$
and $\beta J$, but is approximately independent of the system width $B$. Notice
that square systems may be included in this comparison without problem, as the
extra factor of 2 due to the two possible orientations of the interfaces, is
properly accounted for in the Arrhenius factor.

\item We already indicated repeatedly that our theory may be applied only if
the probability of having more than two interfaces around the cylinder may be
neglected compared to the probability of having just two such interfaces. For
this to be case one has to require $Z_1/Z_0 \ll 1$. For large enough systems 
this condition may be rewritten as $L \ll \exp(\beta\sigma B)$, with $\sigma$ 
the surface tension of the interface. For large systems violation of this 
condition requires extreme aspect ratios, so under normal conditions it will
be satisfied, unless the temperature is very close to the critical one, at
which the surface tension vanishes.
\end{itemize}

We are presently applying the methods described here to a study of nucleation
rates in metastable states. We hope to report on this before long.


\begin{thebibliography}{*99}
\bibitem{mwremark}
McCoy and Wu give the partition function for a system with non-helical boundary
conditions, whereas we have used helical boundary conditions. This should only
make a difference at very low temperatures.
\end{thebibliography}
\end{document}